\documentclass[12pt]{article}
\usepackage{graphicx,epsfig}
\usepackage [latin1]{inputenc}
\usepackage{color}
\title{
Magnetic phase diagram of the Ising model with the long-range 
RKKY interaction}
\author{Lubom\'ira Regeciov\'a and Pavol Farka\v sovsk\'y\\
Institute  of  Experimental  Physics,  Slovak   Academy   of
Sciences\\
Watsonova 47, 040 01 Ko\v {s}ice, Slovakia}
\date{}
\begin{document}
\baselineskip=24pt
\maketitle

\begin{abstract}
The standard Metropolis algorithm and the parallel tempering method are used
to examine magnetization processes in the Ising model with the long-range
RKKY interaction on the Shastry-Sutherland lattice. It is shown that the Ising
model with RKKY interaction exhibits, depending on the value of the Fermi
wave vector $k_F$, the reach spectrum of magnetic solutions, which is manifested in the
appearance of new magnetization plateaus on the magnetization curve. 
In particular, we have found the following set of individual magnetization plateaus 
with fractional magnetization $m/m_s$=1/18, 1/9, 1/8, 1/5, 1/4, 1/3, 3/8, 5/12, 1/2,
3/5, 2/3, which for different values of $k_F$ form various sequences of plateaus,
changing from very complex, appearing near the point $k_F=2\pi/1.2$, to very
simple appearing away this point. Since the change of $k_F$ can be induced 
by doping (the substitution of rare-earth ion by other magnetic ion that 
introduces the additional electrons to the conduction band) the model 
is able to predict the complete sequences of magnetization plateaus, 
which could appear in the tetraboride solid solutions.

\end{abstract}

\newpage
\section{Introduction}

In the past decade, a considerable amount of effort has been devoted 
to understanding the underlying physics that leads to anomalous magnetic properties 
of metallic Shastry-Sutherland  magnets~\cite{Gabani,Matas,Michimura,Yoshii}. 
However, in spite of an impressive research activity, the properties 
of these systems are far from being understood. In particular, this concerns 
the entire group of rare-earth metal tetraborides $RB_4$ ($R=La-Lu$) that exhibits 
the strong geometrical frustration. These compounds have the tetragonal crystal symmetry 
$P4/mbm$ with magnetic ions $R^{3+}$ located on an Archimedean lattice (see Fig.1a)
that is topologically equivalent to the so-called Shastry-Sutherland lattice~\cite{Shastry} 
(see Fig.~1b). 
\begin{figure}[h!]
\begin{center}
\includegraphics[width=7.2cm]{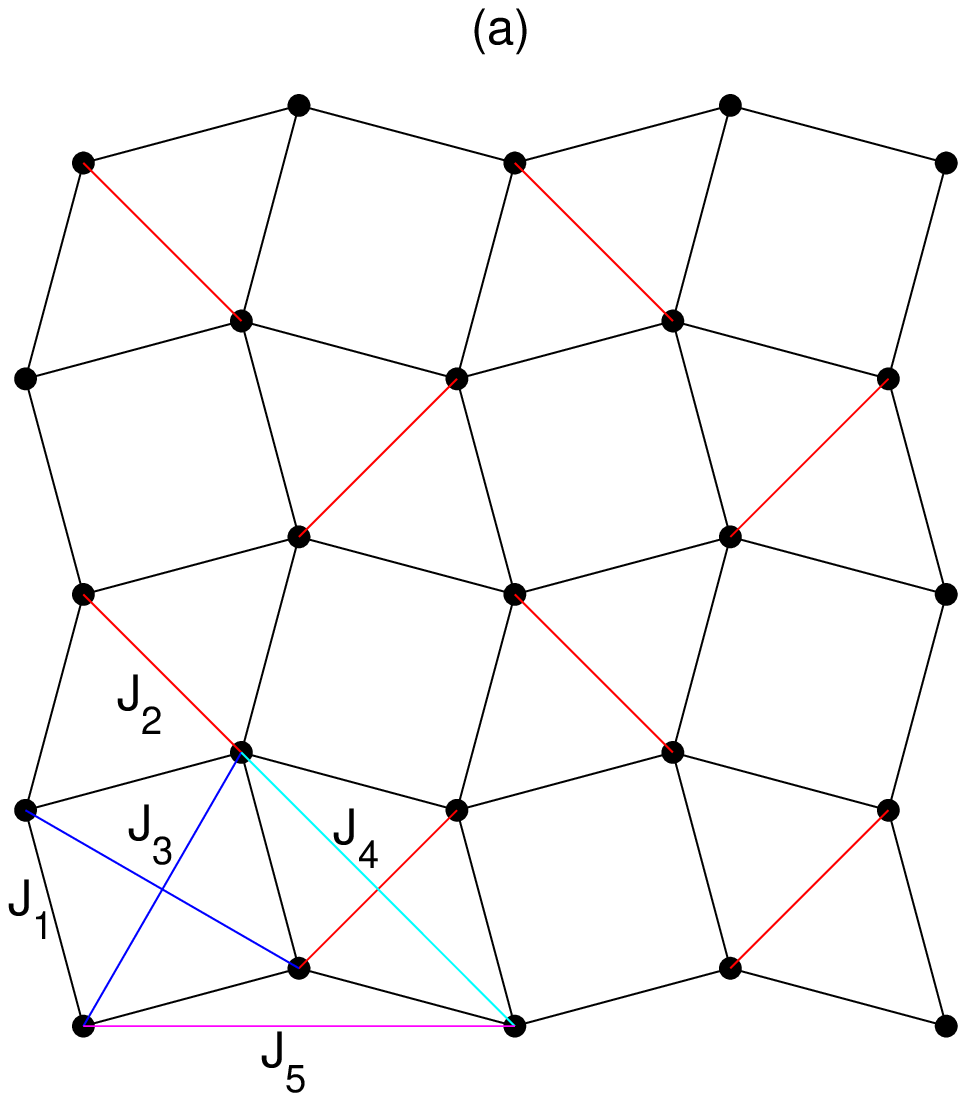}
\includegraphics[width=7.2cm]{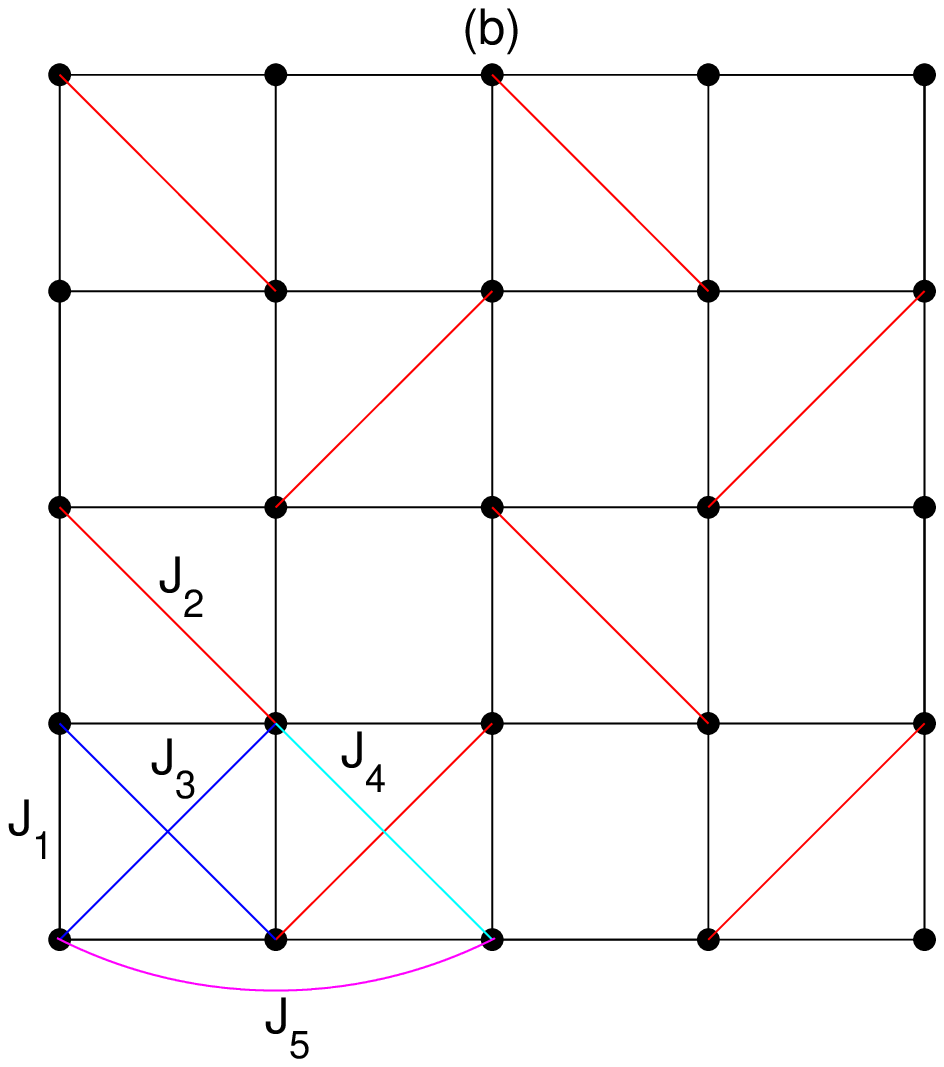}
\end{center}
\caption{\small The real structure realized in the (001) plane 
of rare-earth tetraborides (a), which is topologically identical to the
Shastry-Sutherland lattice (b). $J_1,J_2,J_3,J_4$ and $J_5$ denote
the first, second, third, fourth and fifth nearest neighbors on the 
real Archimedean lattice.    
}
\label{fig1}
\end{figure}
It is supposed that
the anomalous properties of these systems are caused by the geometrical
frustration that leads to an extensive degeneracy in the ground state.
The most famous manifestation of the geometrical frustration in the 
above-mentioned tetraborides is the observation of the fascinating 
sequence of magnetization plateaus with the fractional magnetization. 
For example, for $ErB_4$ the magnetization plateau has been found at 
$m/m_s=1/2$~\cite{Matas,Michimura}, for $TbB_4$ at $m/m_s=$2/9, 1/3, 4/9, 
1/2  and 7/9~\cite{Yoshii}, for $HoB_4$ at $m/m_s=$1/3, 4/9 and 3/5~\cite{Matas} 
and for $TmB_4$ at $m/m_s=$1/11, 1/9, 1/7 and 1/2~\cite{Gabani}.

Despite the metallic nature of rare-earth tetraborides, the first
theoretical works devoted to magnetization processes in these materials 
ignored completely the existence of the conduction electrons, and
exclusively, 
only the spin models have been considered as the generic models for a description
of magnetization plateaus with fractional magnetization. Because of strong
crystal field effects, which are present in rare-earth tetraborides,
the physically reasonable spin model seems to be spin-1/2 Shastry-Sutherland 
model under strong Ising anisotropy~\cite{Gabani}. Thus, the study of
the Ising limit was the first natural step towards the complete understanding of
magnetization processes in rare-earth tetraborides. The subsequent
analytical~\cite{Dublenych1,Dublenych2,Dublenych3,Deviren} 
and numerical~\cite{Chang,Huang,Farky1,Farky2} studies 
showed that the basic version of the Ising model on the Shastry-Sutherland 
lattice, with nearest and next-nearest neighbor interactions
and its extensions, up to the 5th nearest neighbors, are able to describe 
some of the  individual plateaus observed in rare-earth materials, as well
as the partial sequences consisting of two or even three right magnetization
plateaus, but not the complete sequences. These theoretical works point to
the fact that for the correct description of magnetization processes 
in rare-earth tetraborides one has to consider the long-range interactions. 
On the other hand, some theoreticians
speculate that for a description of complete sequences of magnetization
plateaus in rare-earth tetraborides it is necessary to take into account
both, the spin and electron subsystems, as well as the interaction between
them. Indeed, our previous numerical studies showed~\cite{Farky3} that the model based
on the coexistence of both subsystems has a great potential to describe,
at least qualitatively, the complete sequence of magnetization plateaus
observed experimentally in some rare-earth tetraborides, e.g., TmB$_4$.
However, it is questionable if it is the intrinsic property of a model,
or only a consequence of the large number of variables (fitting parameters)
that enter to the model as interaction parameters describing possible spin,
electron and electron spin interactions. 

An alternative model, which takes into account both, the long-range
interactions as well as the presence of conduction electrons, has been
introduced recently by Feng et. al.~\cite{Feng}. Strictly said, it is
a generalized Ising model, in which two spins on lattice sites $i$ and $j$
interact via the RKKY interaction $J_{ij}$ mediated by conduction electrons.
It is supposed that the RKKY coupling between the localized $f$ and 
conduction $s$ electrons is predominant in rare-earth compounds~\cite{Gabani} and 
may play an important role in the mechanism of formation of magnetization
plateaus with fractional magnetizations. The model was studied numerically
and various magnetization plateaus, depending on the value of the Fermi wave
vector $k_F$ of conduction electrons were confirmed. 
However, the importance of these results for a description of magnetization 
processes (magnetization plateaus) in rare-earth tetraborides is questionable, 
since they have been obtained under the assumption that these systems are
electronically three dimensional and the Fermi surface is isotropic, which 
contradicts the latest experimental measurements of the angle-dependent 
magnetotransport in TmB$_4$ revealing the anisotropic Fermi surface 
topology~\cite{Mitra}. For this reason we have decided to examine effects 
of the Fermi surface anisotropy on the magnetic phase diagram (magnetization 
plateaus) of the two-dimensional Ising model with the long range RKKY 
interaction.  For simplicity we consider here only the case of strong 
Fermi-surface anisotropy, represented by the purely electronically two-dimensional 
system, for which the matrix elements of the RKKY interaction have been derived by 
B\'eal-Monod~\cite{Monod} and have the form:
\begin{equation}
J_{ij}=\frac{k_F^2}{2\pi}[B^{(1)}_0(k_Fr_{ij})B^{(2)}_0(k_Fr_{ij})+B^{(1)}_1(k_Fr_{ij})B^{(2)}_1(k_Fr_{ij})],
\end{equation}
where $r_{ij}$ is the distance between the sites $i$ and $j$ on the real 
Archimedean lattice, $k_F$ is the Fermi wave vector, 
$B^{(1)}_n(x)$, with $n=0,1$  are the Bessel functions of the 
first kind and $B^{(2)}_n(x)$ are the Bessel functions of the 
second kind. In the current paper we use this formula for the matrix
elements of the RKKY interaction and construct the comprehensive magnetic
phase diagram of the two-dimensional Ising model with RKKY interaction 
on the Shastry-Sutherland lattice, in which both the magnetic and 
electronic subsystems are considered strictly as two dimensional.
To construct this phase diagram we  
use the same method (the combination of the standard Metropolis 
algorithm and the parallel tempering method) and the same conditions 
(the periodic boundary conditions and the cut-off radius of the RKKY 
interaction $r_{ij}=6$) as were used by Feng et al~\cite{Feng}.

\section{Results and Discussion}
The Hamiltonian of the Ising model with the long-range RKKY interaction 
on the Shastry-Sutherland lattice can be written as
\begin{equation}
H=\sum_{i,j} J_{ij}S^z_iS^z_j - h\sum_i S^z_i\ ,
\end{equation}
where the variable $S^z_i$ denotes the Ising spin with unit length on site $i$, 
$h$ is the magnetic field and the matrix elements $J_{ij}$ 
are given by the formula (2).

To verify the convergence of the Monte-Carlo results we have started our numerical 
studies of the model (3) on the finite cluster of $L=6\times6$ sites, where
also the exact numerical results are accessible. The results of our
numerical calculations are summarized in Fig.~2. The exact magnetization
curve is calculated for T=0, while the Monte-Carlo one is obtained 
for T=0.02 with the $2 \times 10^5$ Monte Carlo steps (the initial 
$1 \times 10^5$ Monte Carlo steps are discarded for equilibrium 
consideration). 
\begin{figure}[h!]
\begin{center}
\includegraphics[width=14.0cm]{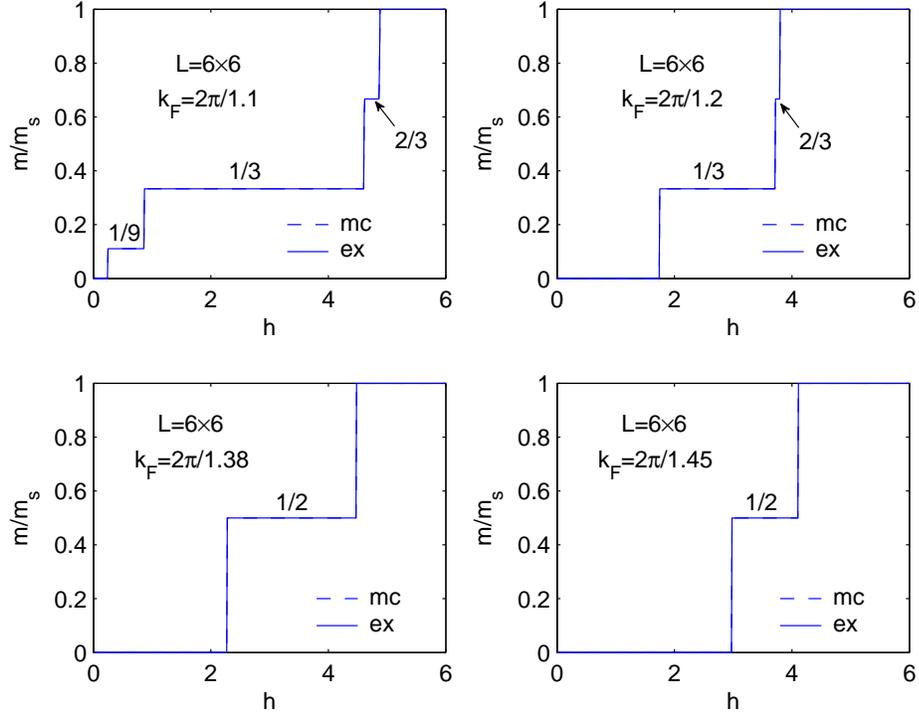}
\end{center}
\caption{\small 
The magnetization curve of the Ising model with RKKY interaction 
calculated for four different values of the Fermi wave vector $k_F$
on the $L=6 \times 6$ cluster by the Monte Carlo method (mc)
and exactly (ex).
}
\label{fig2}
\end{figure}
A comparison of exact and Monte Carlo results shows
that the selected temperature T=0.02 is sufficient to approximate reliably 
the  ground-state properties of the model and that $10^5$ Monte Carlo 
steps (per site) is sufficient to reach well converged results. 
Moreover, these results reveal some interesting physical facts,
concerning the influence of the long-range RKKY interaction on the 
formation of magnetization plateaus, which point to the importance of this 
interaction for a description of real rare-earth tetraborides.
Indeed, our results show that the stability regions of different magnetization 
plateaus with fractional magnetization are very sensitive to the value of
the Fermi wave vector $k_F$, which is directly connected with the concentration 
of conduction electrons $n_e$. However, the change of $n_e$ (and consequently
$k_F$) can be induced  by doping (the substitution of rare-earth ion by other 
magnetic ion that introduces the additional electrons/holes to the conduction band) 
and thus, this simple model could yield the physics for a description of
magnetization processes in doped rare-earth tetraborides. For this reason, we
examine in this paper the comprehensive magnetic phase diagram of the model
in the $k_F-h$ plane and predict the complete sequences of magnetization plateaus
at different $k_F$. To reveal the basic structure of the magnetic phase
diagram we have performed numerical calculations on the finite $L=24 \times 24$ 
cluster (the cluster of the same size has been used also by Feng et
al~\cite{Feng}) and subsequently the exhaustive finite-size scaling analysis of 
the model is done. The results of our Monte-Carlo 
simulations obtained on the $L=24 \times 24$  cluster are summarized in Fig.~3. 
\begin{figure}[h!]
\begin{center}
\includegraphics[width=14.0cm]{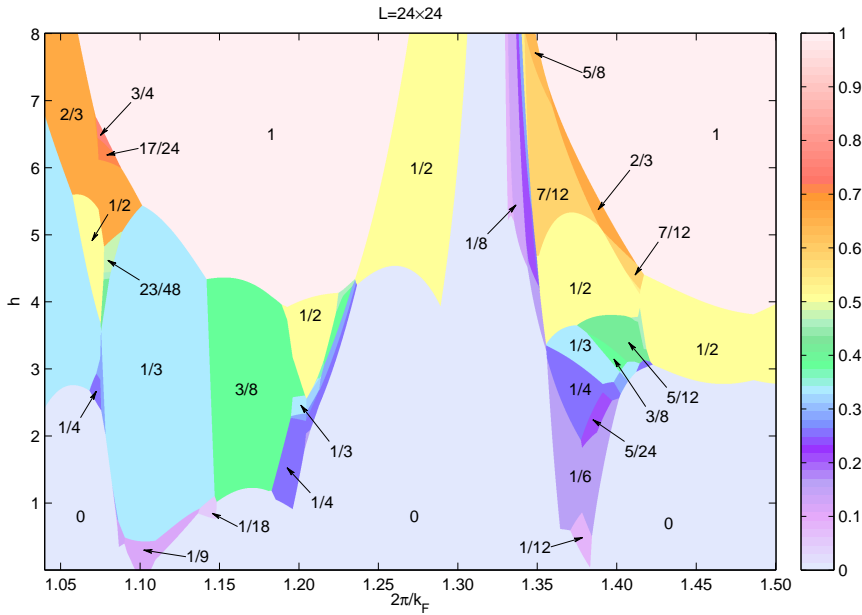}
\end{center}
\caption{\small The magnetic phase diagram of the model in the $k_F-h$
plane calculated for the $L=24 \times 24$ cluster.
}
\label{fig3}
\end{figure}
One can see that the comprehensive magnetic phase diagram of the 
two-dimensional Ising model with the  two-dimensional RKKY interaction 
has much complex structure than the one obtained by Feng et al~\cite{Feng} 
for the three-dimensional one and larger is also the number of magnetization
plateaus. In particular, we have found the following set of individual magnetization 
plateaus with fractional magnetization (only the plateaus with the largest
stability regions are listed):  
$m/m_s$=1/18, 1/12, 1/9, 1/8, 1/6, 1/4, 1/3, 3/8, 5/12, 1/2, 7/12, 2/3,
which for different values of $k_F$ form various sequences of plateaus,
changing from very complex, appearing near the points $k_F=2\pi/1.2$ 
and $k_F=2\pi/1.38$, to relatively simple appearing away these points.
Before discussing these sequences, let us present some interesting 
observations concerning the individual plateaus  with the largest 
stability regions. From all phases corresponding to magnetization plateaus 
with fractional magnetization the largest stability regions exhibit the 
1/3 and 1/2 plateau phases. For the 1/3 plateau phase this 
result is expected since it appears already in the simplest version of 
the Ising model on the Shastry-Sutherland lattice (when only the first ($J_1$)
and second ($J_2$) nearest neighbour interactions are considered), as well
as in practically all extensions of the model, which take into account the next 
nearest neighbor interactions. What is unexpected, however, is the fact that
the 1/3 plateau phase is absent in the central part of the phase diagram
(near the point $k_F=2\pi/1.24$), which corresponds to the real situation 
in undoped rare-earth tetraborides~\cite{Hou}. Thus our results could yield an answer 
on  a very important question concerning the magnetization processes in rare-earth 
tetraborides, and namely, why the 1/3 plateau is absent in these materials, 
while in the conventional Ising model and its next-nearest-neighbour extensions 
(which model very realistically the physical situation in rare-earth tetraborides) 
this plateau practically always exists (and usually it is the largest 
magnetization plateau). The reason seems to be just the long-range RKKY 
interaction, mediated by conduction electrons, which is present in these systems 
and which turns on/off the individual magnetization plateaus according 
to changes of $k_F$. In the central part of the magnetic phase diagram 
(near the $k_F=2\pi/1.24$) the model predicts only the 1/2 plateau, which
perfectly accords with the real situation in the $ErB_4$ 
compound~\cite{Matas,Michimura}, while for $k_F$ slightly smaller than 
$k_F=2\pi/1.24$, it predicts the main 1/2 plateau  accompanied by a sequence 
of narrow magnetization plateaus with $m/m_s<1/2$, similarly as was observed 
in $TmB_4$ compound~\cite{Gabani}.
Unfortunately, the size of cluster used in our calculations consisting 
of $L=24 \times 24$ sites is still small to verify the magic sequence 
of magnetization plateaus observed in $TmB_4$ consisting of 
$m/m_s$=1/2, 1/7, 1/9, and 1/11 plateaus (to verify this, the cluster of 
size at least $L=1386 \times 1386$ should be considered), but it seems 
that the Ising model with long-range Coulomb interaction could yield, 
at least qualitatively, the correct physics to describe magnetization 
processes in these materials. 

Untill now we have discussed mainly the effects of the long-range
RKKY interaction on the formation of different magnetization plateaus,
without a deeper interest, which types of spin arrangements stand
behind these magnetization plateaus. Performing a more detailed analysis
of this problem we have obtained an interesting result, and namely, that
some magnetization plateaus are formed by two, three, or even more different
spin configurations. In particular, we have found 
(i) four different spin configurations corresponding to the 1/2 magnetization 
plateau, (ii) five different spin configurations corresponding to the 1/3 
magnetization plateau, and (iii) eight different spin configurations 
corresponding to the zero magnetization plateau. The complete list of spin 
configurations corresponding to these magnetization plateaus with the largest
stability regions in the magnetic phase diagram are displayed in Fig.~4. 
\begin{figure}[h!]
\begin{center}
\includegraphics[width=14.0cm]{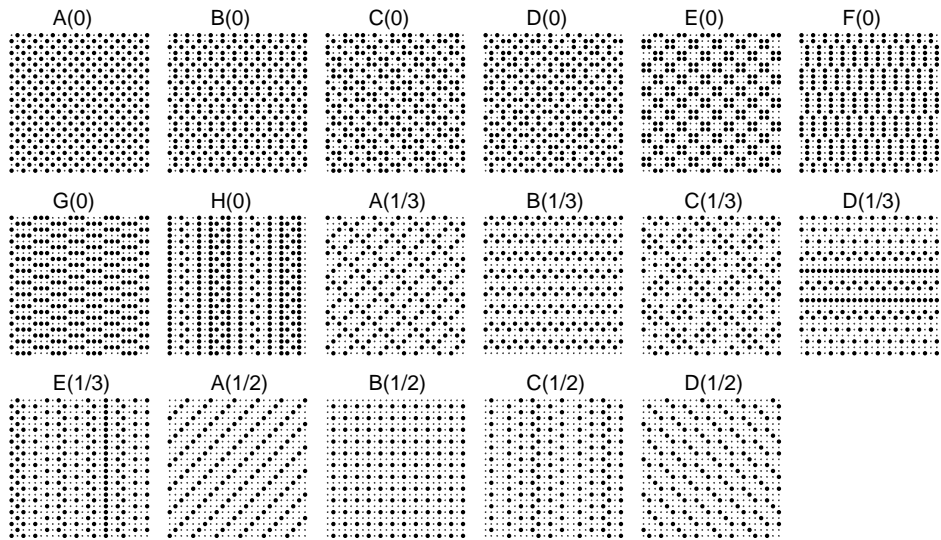}
\end{center}
\vspace{-1cm}
\caption{\small The complete list of spin configurations corresponding to magnetization plateaus with 
the largest stability regions ($m/m_s$=0, 1/3, 1/2) from Fig.~3.
}
\label{fig4}
\end{figure}
One can see that the spectrum of
configuration types is very wide and includes various types of axial and
diagonal striped phases as well as homogeneous (quasi-homogeneous)
distributions of single spins or $n$-spin clusters, confirming  strong 
influence of the long-range RKKY interaction  on the ground state 
properties of the model.

Of course, one can object that these results can not be considered as definite
since they were obtained on the relatively small cluster consisting of only 
$L=24 \times 24$ sites. For this reason we  have performed exhaustive 
numerical studies of the model on much larger cluster consisting
of $L=120 \times 120$. Numerical calculations on such a large cluster 
are however extremely time-consuming and therefore we had to decrease
the number of Monte-Carlo steps from $2 \times 10^5$ (the case of 
$L=24 \times 24$ cluster) to $2 \times 10^4$, but it seems that 
this fact did not influence significantly the convergence of Monte-Carlo
results (see Fig,~5). 
\begin{figure}[h!]
\begin{center}
\includegraphics[width=14.0cm]{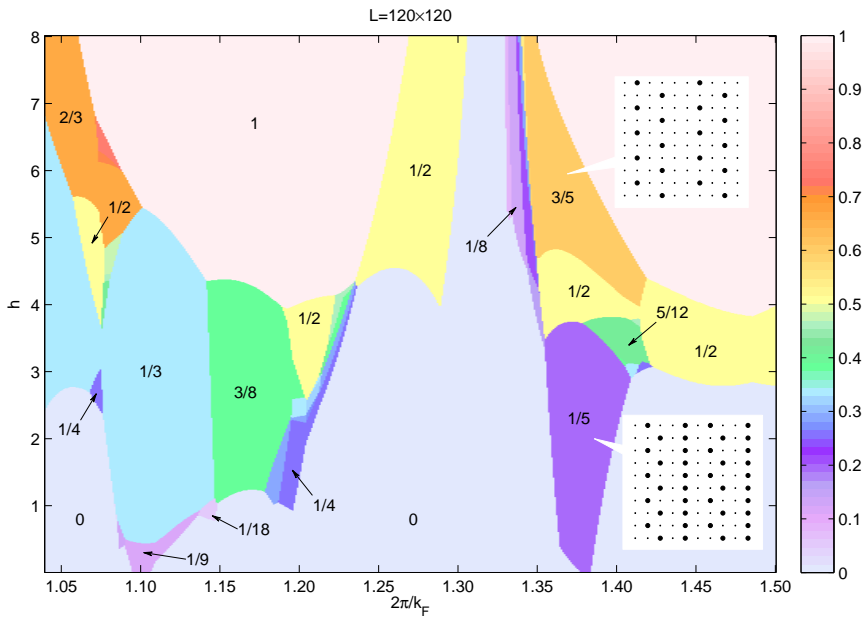}
\end{center}
\caption{\small The magnetic phase diagram of the model in the $k_F-h$
plane calculated for the $L=120 \times 120$ cluster.
}
\label{fig5}
\end{figure}
Indeed, a direct comparison of magnetic phase diagrams obtained
on $L=24 \times 24$ and $L=120 \times 120$ cluster shows that they are
practically identical for $k_F<2\pi/1.3$ and small differences are observed 
only for $k_F > 2\pi/1.3$, strictly said near the point $k_F=2\pi/1.38$. In this
region the magnetic phase diagram of the model obtained on $L=24 \times 24$
cluster exhibits relatively complex structure formed by the main 1/6
plateau accompanied by smaller 1/4, 5/24 and 1/12 plateaus, while 
for the $L=120 \times 120$ cluster there exists only one large 1/5 
plateau. The corresponding plateau phase has the period 10 and therefore
it can not appear in the $L=24 \times 24$ magnetic phase diagram, but was
replaced by phases with nearest fractional magnetizations. For the same 
reason the 7/12 plateau phase is replaced by the 3/5 plateau phase. Since 
the $L=120 \times 120$ cluster is indeed the robust one and no significant 
finite-size effects have been observed comparing results obtained for $L=24 \times
24$ and $L=120 \times 120$, we suppose that the results presented in Fig.~5
can be used satisfactorily for a description of macroscopic systems.
This conjecture is supported also by our additional results presented 
in Fig.~6, where the ground-state energies (per site) of the model calculated 
on two different clusters of $L=120 \times 120$ and $L=140 \times 140$ sites 
are compared. The cluster of $L=140 \times 140$ has been chosen for 
the reason that it is compatible with the 1/7 plateau, which absent on 
the $L=120 \times 120$ cluster.
\begin{figure}[h!]
\begin{center}
\includegraphics[width=10.0cm]{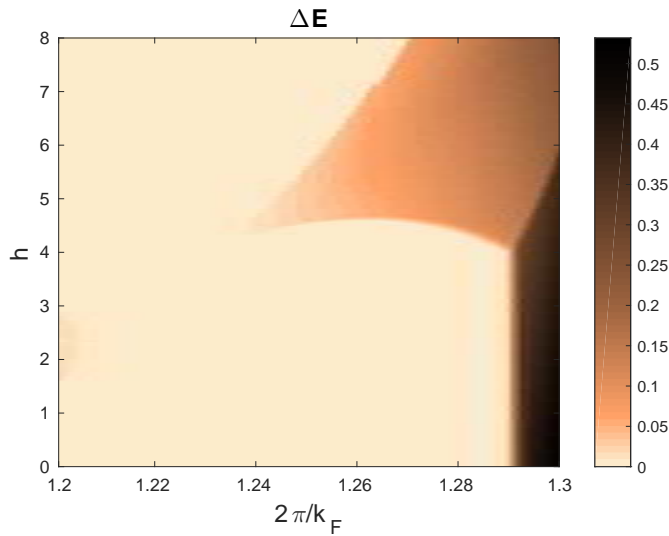}
\end{center}
\caption{\small The difference $\Delta E=E_{140}-E_{120}$ between the ground-state 
energies (per site) of the model calculated on the $L=120 \times 120$ and 
$L=140 \times 140$ cluster.}
\label{fig6}
\end{figure}
One can see that for all values of the Fermi wave vector $k_F$ from the central 
part of the phase diagram (which corresponds to the real situation in undoped 
rare-earth tetraborides~\cite{Hou}) the ground-state energy of the model on the 
$L=120 \times 120$ cluster is smaller than one corresponding to the 
$L=140 \times 140$ cluster and thus no new plateaus compatible with the 
$L=140 \times 140$ cluster, e.g. the 1/7 plateau, are expected in the 
thermodynamic limit $L \to \infty$.   

One of possible explanations of absence of the 1/7 plateau in our
results could be the fact that the 2D formula used in our work for the 
matrix elements $J_{ij}$ of the RKKY interaction is too crude approximation 
of the real situation in TmB$_4$, where the Fermi surface is anisotropic, 
but not 2D. Therefore, in a more realistic model one should consider, 
instead the 2D~\cite{Monod} or 3D~\cite{Feng} formula,  an intermediate 
version of the RKKY interaction between 2D and 3D. It is  also possible 
that the model based on the indirect interaction between  two spins 
mediated by conduction electrons is insufficient to capture all
experimentally observed features of magnetization processes in rare-earth 
tetraborides, which are metallic and for their correct description 
it will be necessary to use a more complex model taking into account both, 
the electron and spin subsystems as well as direct interactions within 
subsystems and between them. Works in both directions, the generalization 
of the RKKY interaction for the case of the anisotropic Fermi surface and 
numerical studies within a more complex model, are currently in progress.

In Table~1 we present several sequences of magnetization plateaus 
that follow from our phase diagram for selected values of the Fermi wave 
vector $k_F$  representing typical behaviours of the model from the 
regions $k_F<2\pi/1.24$ and $k_F>2\pi/1.24$. 
\begin{table}[h!]
\begin{center}   
\begin{tabular}{|l|l|}
\hline\hline
$2\pi/k_F$ & $m/m_s$  \\
\hline
   $1.05$ & $0,1/3,2/3$  \\
   $1.07$ & $0,1/3,1/2,2/3,1$  \\
   $1.095$ & $0,1/9,1/3,2/3,1$  \\
   $1.12$ & $0,1/9,1/3,1$  \\
   $1.17$ & $0,3/8,1$  \\
   $1.196$ & $0,1/4,3/8,1/2,1$  \\
   $1.25$ & $0,1/2,1$  \\
   $1.37$ & $0,1/5,1/2,3/5,1$  \\
   $1.4$ & $0,1/5,5/12,1/2,3/5,1$  \\
   $1.45$ & $0,1/2,1$  \\
\hline\hline
\end{tabular}
\end{center} 
\vspace*{-0.5cm}
\caption{\small Representative sequences of magnetization plateaus
identified at different values of $k_F$.}
\label{tab6}
\end{table}
From the point of view of rare-earth 
tetraborides (e.g., $TmB_4$,$ErB_4$), the set of $k_F$ with $k_F<2\pi/1.24$ 
models quite realistically the experiment when the additional holes are 
doped into the system, while the second one with $k_F>2\pi/1.24$ models 
the experiment, when the additional electrons are doped into the system. 
It is seen that the spectrum of magnetization sequences is very wide 
and thus the results obtained can serve as a motivation for experimental studies 
of the influence of doping on the formation of magnetization plateaus 
in the tetraboride solid solutions.

In summary, we have presented a simple model for a description of magnetization processes 
in metallic rare-earth tetraborides. It is based on the two-dimensional Ising model, 
in which two spins on the Shastry-Sutherland lattice interact via the long-range RKKY 
interaction $J_{ij}$ mediated by conduction electrons. The model is solved 
by a combination of the standard Metropolis algorithm and the parallel tempering 
method and it yields the reach spectrum of magnetic solutions (magnetization
plateaus), depending on the value of the Fermi wave vector $k_F$ and the external 
magnetic field $h$. In particular, we have found the following set of individual 
magnetization plateaus with fractional magnetization 
$m/m_s$=1/18, 1/9, 1/8, 1/5, 1/4, 1/3, 3/8, 5/12, 1/2, 3/5, 2/3, which for different 
values of $k_F$ form various sequences of plateaus, changing from very complex, 
appearing near the points $k_F=2\pi/1.2$  to relatively simple appearing away 
this point. Since the change of $k_F$ can be induced by doping (the substitution 
of rare-earth ion by other magnetic ion that introduces the additional electrons 
(holes) into the system) the model is able to predict the complete sequences 
of magnetization plateaus, that could appear in the tetraboride solid solutions.

\vspace{0.5cm}
{\small This work was supported by projects ITMS 26220120047, VEGA 2-0112-18
and APVV-17-0020. Calculations were performed in the Computing Centre 
of the Slovak Academy of Sciences using the supercomputing infrastructure 
acquired in project ITMS 26230120002 and 26210120002 (Slovak infrastructure 
for high-performance computing) supported by the Research and Development
Operational Programme funded by the ERDF.}


\newpage


\begin{thebibliography}{[1]}
\bibitem{Gabani}  
K. Siemensmeyer, E. Wulf, H.~J. Mikeska, K. Flachbart, S. Gabani, S. Matas, P. Priputen, A. Efdokimova, 
and N. Shitsevalova, Phys. Rev. Lett. {\bf 101}, 177201 (2008).
\bibitem{Matas}
S. Mata\v s, K. Siemensmeyer, E. Wheeler, E. Wulf, R. Beyer, Th. Hermannsdörfer, O. Ignatchik, M. Uhlarz, 
K. Flachbart, S. Gab\'ani, P. Priputen, A. Efdokimova, and N. Shitsevalova, J. Phys. Conf. Ser. {\bf 200}, 032041 (2010).
\bibitem{Michimura}
S. Michimura, A. Shigekawa, F. Iga, M. Sera, T. Takabatake, K. Ohoyama, and Y. Okabe, Physica B  {\bf 378}, 596 (2006).
\bibitem{Yoshii}
S. Yoshii, T. Yamamoto, M. Hagiwara, S. Michimura, A. Shigekawa, F. Iga, T. Takabatake, and K. Kindo, Phys. Rev. Lett. {\bf 101}, 087202 (2008).
\bibitem{Shastry}%
 B.\,S.~Shastry, B.~Sutherland, Physica B  and C \textbf{108}, 1069 (1981).
 \bibitem{Dublenych1}
 Y. Dublenych, Phys. Rev. Lett. {\bf 109}, 167202 (2012).
 \bibitem{Dublenych2}
 Y. Dublenych, Phys. Rev. E {\bf 88}, 022111 (2013).
\bibitem{Dublenych3}
{Y. Dublenych,  Phys. Rev. E {\bf 90}, 052123 (2014).}
\bibitem{Deviren}
{S.A. Deviren, JMMM {\bf 393}, 508 (2015).} 
 \bibitem{Chang}%
 M.\,C.~Chang and M.\,F.~Yang, Phys. Rev. B {\bf 79}, 104411 (2009).
\bibitem{Huang}
W.~C. Huang, L. Huo, J.~J. Feng, Z.~B. Yan, X.~T. Jia, X.~S. Gao, M.~H. Qin, and J.-M. Liu, EPL {\bf 102}, 37005 (2013).
\bibitem{Farky1}
H.~\v{C}en\v{c}arikov\'a and P. Farka\v{s}ovsk\'y,  Phys. Status Solidi B {\bf
252}, 333 (2015).
\bibitem{Farky2}
P. Farka\v{s}ovsk\'y and  L. Regeciov\'a, Eur. Phys. J. B {\bf 92}. 33 (2019).
\bibitem{Farky3}
P. Farka\v{s}ovsk\'y, H.~\v{C}en\v{c}arikov\'a, S. Mata\v s, Phys. Rev. B {\bf 82}, 54410 (2010).
\bibitem{Feng} 
J.~J. Feng, L. Huo, W.~C. Huang, Y. Wang, M.~H. Qin, J.-M. Liu, and Z. Ren, EPL {\bf 105}, 17009 (2014).
\bibitem{Mitra}S. Mitra, J.~G.~S. Kang, J. Shin, J.~ Q. Ng, S.~S. Sunku,
T. Kong, P.~C. Canfield, B.~S. Shastry, P. Sengupta, and Ch. Panagopoulos, Phys. Rev. B {\bf
99}, 045119 (2019).
\bibitem{Monod}M.T. B\'eal-Monod, Phys. Rev. B {\bf 36}, 88835 (1987).
\bibitem{Hou} B.~H. Hou, F.~Y. Liu, B. Jiao,and M. Yue, Acta Phys. Sin. {\bf 61}, 077302 (2012).



\end{thebibliography}
\end{document}